\documentclass[aps,prb, showpacs, reprint, preprintnumbers,amsmath,amssymb,superscriptaddress,floatfix]{revtex4-1}

\usepackage[]{graphicx}      
\usepackage{bm}             	
\usepackage{times}			
\usepackage{tabularx}
\usepackage{color}
\usepackage{dcolumn}		
\usepackage{amsmath}
\usepackage{amssymb}
\usepackage{amsfonts}
\usepackage{times}
\usepackage{tabularx}
\usepackage{xcolor,colortbl}

\usepackage{amsmath}

\makeatletter
\newcommand\xleftrightarrow[2][]{%
  \ext@arrow 9999{\longleftrightarrowfill@}{#1}{#2}}
\newcommand\longleftrightarrowfill@{%
  \arrowfill@\leftarrow\relbar\rightarrow}
\makeatother

\begin{document}
\title{Back-hopping in Spin-Transfer-Torque switching of perpendicularly magnetized tunnel junctions}
\author{T. Devolder}
\email{thibaut.devolder@u-psud.fr}
\affiliation{Universit\'e Paris-Saclay, CNRS, Centre de Nanosciences et de Nanotechnologies, Palaiseau, France}
\author{O. Bultynck}
\affiliation{KU Leuven, Department of Materials Engineering, Kasteelpark Arenberg 44, 3001 Leuven, Belgium}
\affiliation{imec, Kapeldreef 75, 3001 Heverlee, Belgium}
\author{P. Bouquin}
\affiliation{Universit\'e Paris-Saclay, CNRS, Centre de Nanosciences et de Nanotechnologies, Palaiseau, France}
\affiliation{imec, Kapeldreef 75, 3001 Heverlee, Belgium}
\author{V. D. Nguyen}
\author{S. Rao}
\author{D. Wan}
\affiliation{imec, Kapeldreef 75, 3001 Heverlee, Belgium}
\author{B. Sor\'ee}
\affiliation{imec, Kapeldreef 75, 3001 Heverlee, Belgium}
\affiliation{KU Leuven, Department of Electrical Engineering, Kasteelpark Arenberg 10, 3001 Leuven, Belgium}
\affiliation{University of Antwerp, Physics Department, Groenenborgerlaan 171, B-2020 Antwerpen, Belgium}
\author{I. P. Radu}
\author{G. S. Kar}
\author{S. Couet}
\affiliation{imec, Kapeldreef 75, 3001 Heverlee, Belgium}

\date{\today}                                           
%
%
\begin{abstract}
We analyse the phenomenon of back-hopping in spin-torque induced switching of the magnetization in perpendicularly magnetized tunnel junctions. The analysis is based on single-shot time-resolved conductance measurements of the pulse-induced back-hopping. 
Studying several material variants reveals that the back-hopping is a feature of the nominally fixed system of the tunnel junction. 
The back-hopping is found to proceed by two sequential switching events that lead to a final state P' of conductance close to --but distinct from-- that of the conventional parallel state. The P' state does not exist at remanence. It generally relaxes to the conventional antiparallel state if the current is removed. The P' state involves a switching of the sole spin-polarizing part of the fixed layers. The analysis of literature indicates that back-hopping occurs only when the spin-polarizing layer is too weakly coupled to the rest of the fixed system, which justifies a posteriori the mitigation strategies of back-hopping that were implemented empirically in spin-transfer-torque magnetic random access memories. 
\end{abstract}

\maketitle

%
%

\section{Introduction}
The mechanism of magnetization switching is a central problem of the magnetism community. Its understanding in nanosized ultrathin systems has become of practical interest thanks to the emergence of spin transfer torque (STT) magnetic random access memories (STTMRAM) \cite{khvalkovskiy_basic_2013}. This technology is based on Magnetic Tunnel Junctions (MTJ) consisting of a so-called free layer (FL) whose magnetization can be switched though the transfer of spins from --or to-- a complex stack made of a succession of layers with nominally fixed magnetizations. The MTJ resistance is most often low (resp. high) in the so-called Parallel (P) (resp. antiparallel, AP) when the FL magnetization is parallel (resp. antiparallel) to that of the closest layer of the fixed system. In addition to its fundamental interest, understanding how reliably and fast STT can set the resistance state of an MTJ is of critical importance \cite{hu_reliable_2019} for magnetic memory technologies.

A major obstacle to reliable and fast STT-induced switching is a counterintuitive phenomenon: the back-hopping\cite{min_back-hopping_2009} (BH). We generally expect that the switching probability should always increase with the applied voltage. However the contrary can sometimes occur: in many instances, the MTJ resistance can back-hop to its original state after the apparent successful switching of the FL magnetization \cite{min_back-hopping_2009, sun_high-bias_2009}. 
In the initial experiments done on in-plane magnetized junctions, the BH was much more severe for the AP to P transition. As this corresponds to when the electrons tunnel into the FL, it was conjectured \cite{sun_high-bias_2009} to result from hot-electron processes altering the STT inside the FL. This idea of a transport-based origin of BH was rationalized later by Skowro\'nski et al. \cite{skowronski_backhopping_2013, frankowski_backhopping_2015}. They proposed that the BH could result from a combination of the Slonczewski torque and of the field-like spin-torque acting on the free layer at high bias. 
Theodonis et al. indeed showed \cite{theodonis_anomalous_2006} that in some MTJs there is a reversal of the sign of the Slonczewski torque at large voltage bias, potentially inducing the back reversal of the free layer, especially if complemented by potentially existing field-like torque. Till then, the BH phenomenon was tacitly believed \cite{min_back-hopping_2009, sun_high-bias_2009, skowronski_backhopping_2013, frankowski_backhopping_2015} to be a consequence of the sole FL dynamics. 

An alternative explanation was proposed more recently from time-resolved characterizations \cite{devolder_time-resolved_2016,safranski_interface_2019} or short pulse-induced characterizations \cite{hu_reliable_2019} of the BH phenomenon on systems with perpendicular magnetic anisotropy (PMA). These measurements suggested that the BH was a dynamical process involving magnetic configurations in which the magnetizations of the fixed layers were \textit{also} destabilized at high voltage biases. If the voltage was switched off while part of the fixed layers is in an unconventional state (i.e. with a layer magnetized in a direction distinct from the nominal design), the device could relax to the non-targeted state, yielding a high write error rate. The conclusion was that it was not legitimate to neglect the spin-torque acting on the fixed layers. This was modeled soon after by Abert et al. \cite{abert_back-hopping_2018} who developed a spin-diffusion model to enable an accurate description of the torques in a simplified MTJ. 
With these accurate torques, high bias was predicted to lead to undesired switching of the fixed layer, which then induces fast perpetual cyclic switching of both the FL and pinned layers, as indeed compatible with some experimental observations \cite{thomas_spin_2017}. 

This survey of the studies of back-hopping indicates two proposed origins of the phenomenon: the bias dependence of the torques acting on the FL \cite{sun_high-bias_2009, skowronski_backhopping_2013, frankowski_backhopping_2015}, or the dynamics of the reference layer\cite{devolder_time-resolved_2016, thomas_spin_2017, abert_back-hopping_2018, safranski_interface_2019}. In this work, we discriminate between these two scenarios by measuring in a single-shot time-resolved manner the BH in several MTJs having different FLs but sharing a fixed system with layers of identical nominal compositions. 
The  time-resolution allows to identify the intermediate states during the BH. We find in particular that the BH proceeds by two successive switching steps leading to a final state of conductance close to --but distinct from-- that of the P state. We show that this P' state involves a switching of the sole spin-polarizing part of the fixed layers, and it occurs when the spin-polarizing layer is too weakly coupled to the rest of the fixed system. This sheds light onto the mitigation strategies of BH to implement in advanced STT-MRAM stacks with minimal high bias write error rates. 

\section{Samples and methods} 
\begin{figure}
\hspace*{-0 cm}\includegraphics[width=8.5 cm]{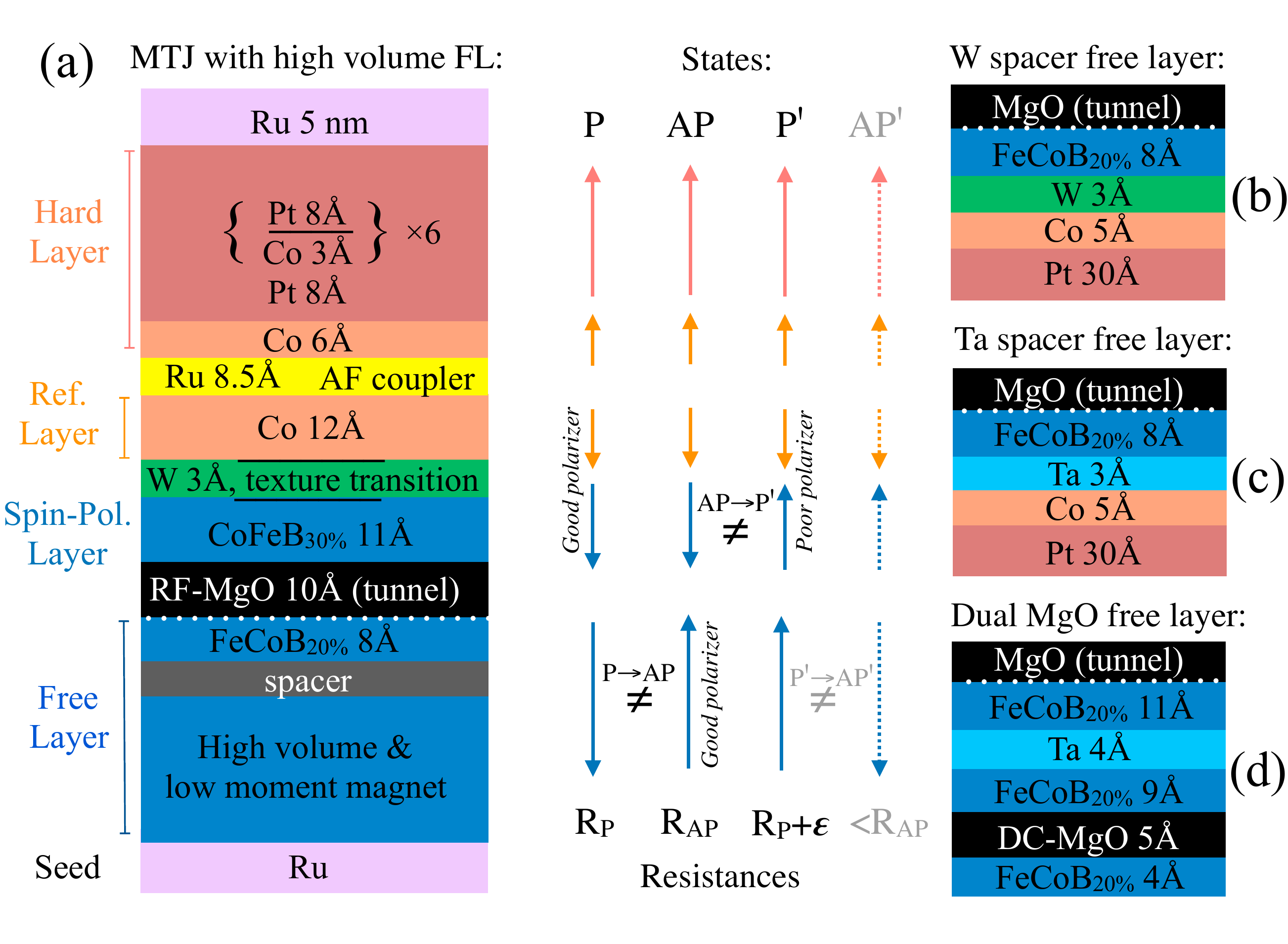}{\centering}
\caption{\textcolor{black}{Nominal sample compositions and definition of the magnetization configurations. (a) Full MTJ stack with a high volume low moment free layer}. The vertical arrows sketch the magnetizations of the different functional magnetic blocks within the full stack. (b) Hybrid free layer with W interlayer. (c) Hybrid free layer with Ta interlayer. (d) Dual MgO free layer. }
\label{STACK}
\end{figure}

Over the years, different generations of PMA-MTJ stacks of been studied: some showed clear BH \cite{devolder_time-resolved_2016, devolder_ferromagnetic_2016, safranski_interface_2019}, some others did not \cite{hahn_time-resolved_2016, devolder_annealing_2017, bultynck_instant-spin_2018, devolder_effect_2019}. Since our objective is to understand the phenomenon of BH, we have selected the MTJ stacks where BH is clearly happening. We thus consider the top-pinned MTJs sketched in Fig.~\ref{STACK}, deposited by physical vapor deposition and then annealed at 300$^\circ$C. The fixed system above the MgO tunnel oxide is common to all samples. From bottom to top, the fixed system comprises a FeCoB 11\r{A} spin-polarizing layer (SPL) with body-centered cubic (bcc) structure. It is ferromagnetically coupled to a Co 12\r{A} reference layer (RL) of face-centered cubic (fcc) structure through the W 3\r{A} texture-transition layer. The RL is coupled antiferromagnetically to a high anisotropy fcc [Co/Pt] hard multilayer (HL) though a Ru spacer. 

%

Within these MTJs, all FL always have at least 8\r{A} of FeCoB in contact with the MgO tunnel oxide to warrant a decent TMR, typically around 80\% depending on the devices. We have then varied the other parts of the FL to cover material options leading most probably to substantially different bias dependences of the spin-torques acting on the FL. The standard FL is a dual-MgO FL [Fig.~\ref{STACK}(d)] with full bcc character. 
We study also high anisotropy versions of the FL, in which interface anisotropies from Co/Pt interfaces are harnessed at the cost of bcc-to-fcc texture transition layers [W or Ta, Fig.~\ref{STACK}(b-c)] at the midst of the FL. We finally also consider a proprietary high volume low moment free layer that shows higher thermal stability and better immunity to the stray fields emanating from the fixed system [Fig.~\ref{STACK}(a)].

Although the 4 MTJ have nominally identical tunnel barriers and nearby electrodes, the tiny differences in the material growth affect the resistance-area products that range from 7.8 (dual MgO FL), 9.6 (high volume low moment FL), 14.6 (Ta spacer FL) to $19.9 ~\Omega.\mu\textrm m^2$ (W spacer FL). Each MTJ was patterned into disk-shaped devices of various diameters. During the resistance versus dc voltage loops (Fig.~\ref{RVloops}), the back-side of the samples are held at constant "applied" temperature and the system is in vacuum.

The devices with 60-80 nm diameter appear to be the most adequate for BH studies: they are wide enough to pass the large currents needed for precise electrical measurements and they are found empirically small enough not to exhibit the complexity associated with domain wall pinning during magnetization reversal. 

\section{Results}
\begin{figure}
\hspace*{-0.2 cm}\includegraphics[width=8.5 cm]{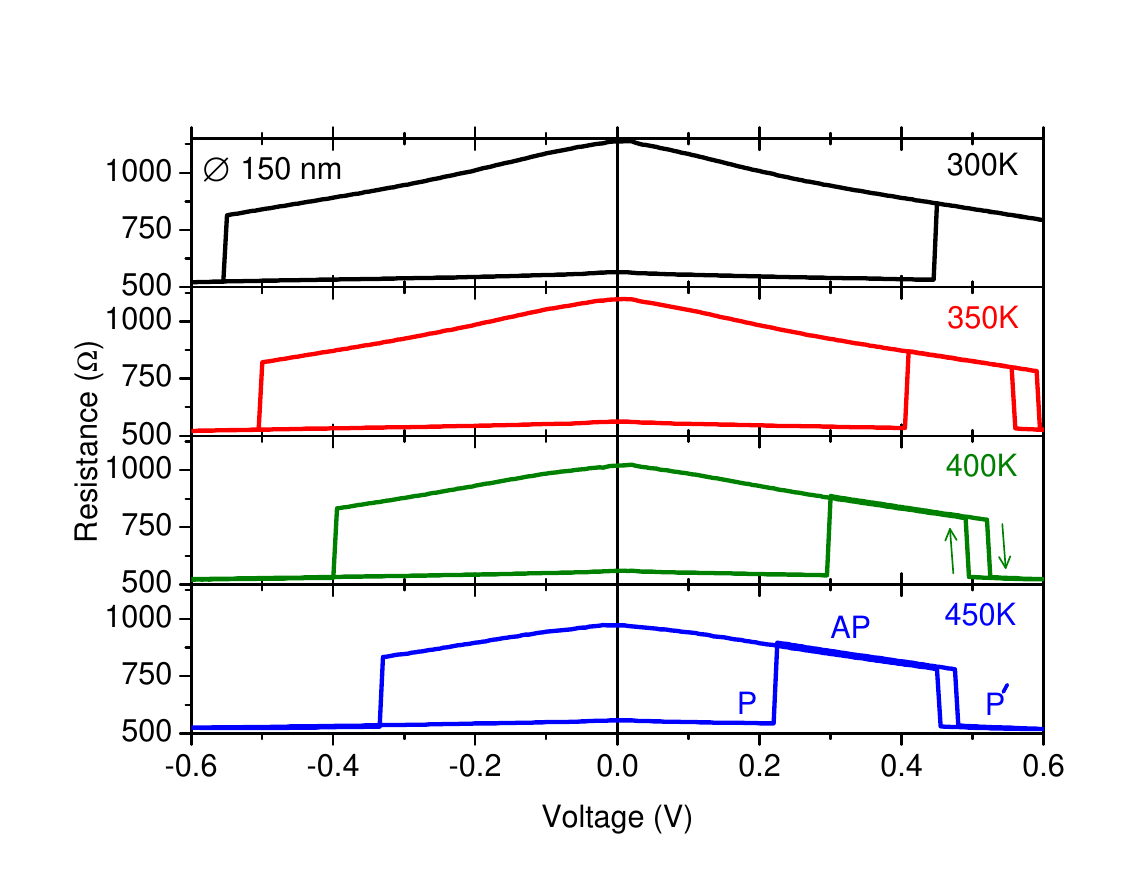}
\caption{(Color online). Hysteresis loops of the resistance versus applied dc voltage for a 150 nm 
diameter MTJ with a high volume low moment free layer. The applied temperatures are (a) 300 K, (b) 350 K, (c) 400 K and (d) 450 K. The labels P, AP and P' define the resistance of the Parallel, Antiparallel and Parallel prime states.}
\label{RVloops}
\end{figure}

The quasi-static consequence of BH is best illustrated in Resistance versus Voltage loops, as in the example of high volume low moment FL devices in Fig.~\ref{RVloops}. When staying below $\pm600~\textrm{mV}$ at room temperature, the devices simply switch back and forth between the P and AP states, thereby forming a conventional STT loop. Higher dc voltage --sometimes damaging the devices-- are needed to reveal the BH at room temperature. Alternatively, BH can also be revealed by heating the devices [Fig.~\ref{RVloops}(b-d)] to reduce all the switching thresholds and place them at safe voltage levels. Above 350K a new state appears at positive bias. It is visited in an hysteretic manner for the high volume low moment FL [see arrows in Fig.~\ref{RVloops}(c)] and for the Ta-based hybrid FL. For the W-based hybrid free layer and the dual MgO FL, the P' state is obtained in an anhysteretic manner, which looks abrupt and reversible at the quasi-static time scale of the $R(V)$ loops. We shall refer to this new state as the P' state as its resistance is close to that of the remanent P state. Note that even if the resistance of the P state is almost independent from the temperature (we have indeed $R_P^{450~\textrm K}=R_P^{300~\textrm K} / 0.982$ at $V_\textrm{applied}=0$) and much less dependent on the bias voltage than the AP state (see Fig.~\ref{RVloops}), the STT loops are not sufficient to determine whether the P and P' states differ or not. They might have been confused in some earlier studies. 
%

\begin{figure}
\includegraphics[width=8.5 cm]{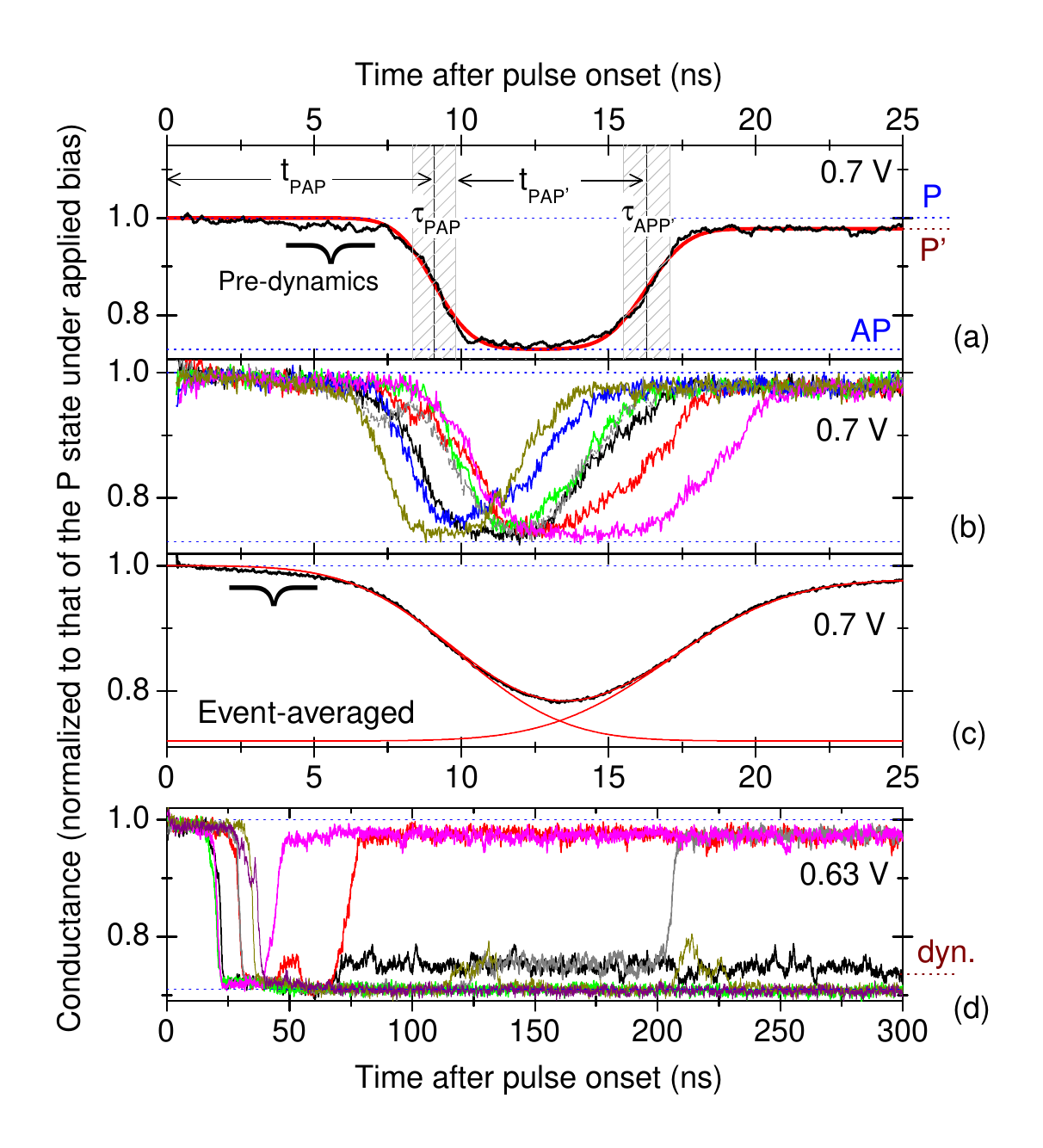}
\caption{Anthology of the back-hopping phenomenon in a 78 nm diameter MTJ with a high volume low moment free layer. The time-resolved curves show the conductance of the MTJ in response to voltage steps with constant plateaus. The smooth lines are fits with sums of error functions forced to match the resistance of the P, AP and P' states. (a) Representative single shot response with 3 GHz bandwidth with the definition of the P$\rightarrow$AP and AP$\rightarrow$P' switching times and their transition times $\tau_\textrm{PAP}$ and $\tau_\textrm{APP'}$. (b) Randomly selected responses recorded with a larger (16.8 GHz) bandwidth. (c) Averaging of (b) over 512 events. (d) Randomly selected single shot responses recorded with 3 GHz bandwidth at a lower applied voltage. Note the transient existence of a noisy (labeled 'dyn') of intermediate conductance.}
\label{ANTHOLOGY}
\end{figure}

\begin{figure}
\hspace*{-0.5 cm}\includegraphics[width=8.5 cm]{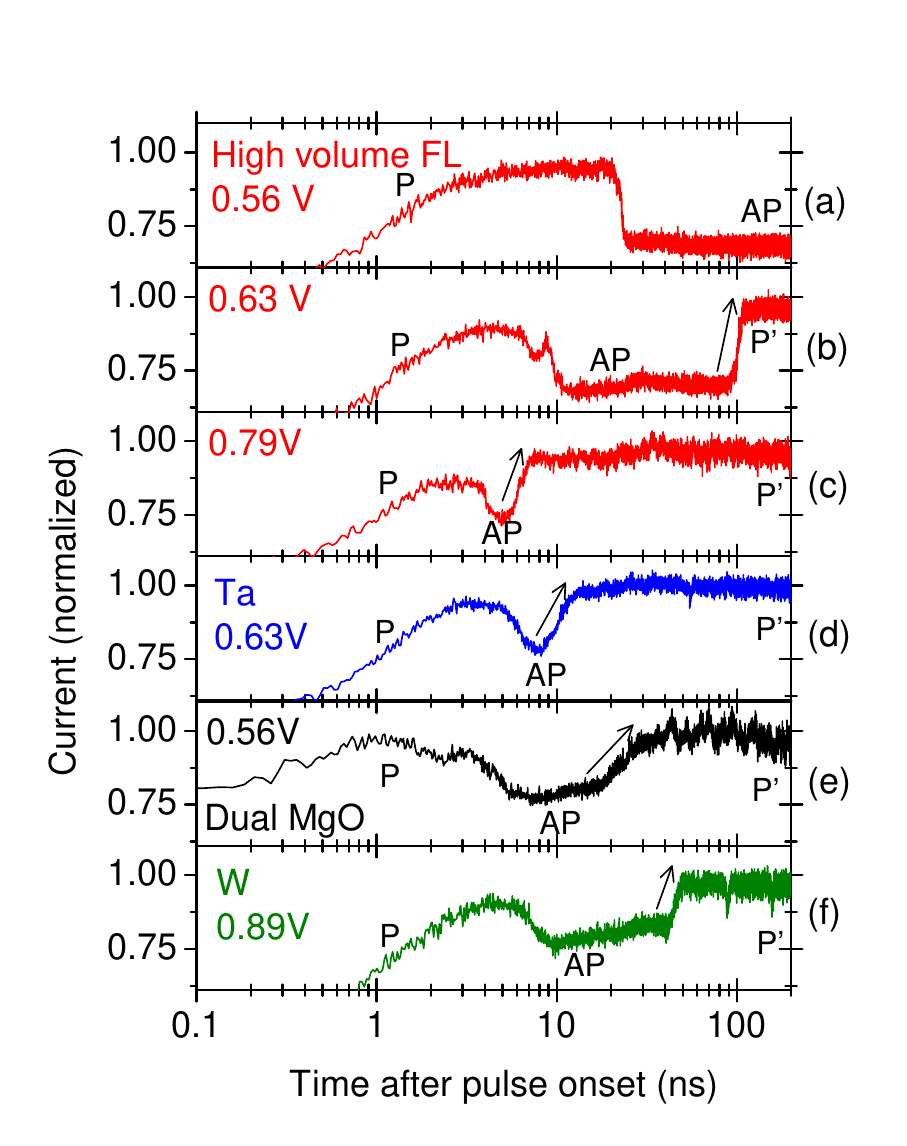} 
\centering
\caption{Time-resolved back-hopping in 90 nm disks made from the MTJs of Fig.1. (a) absence of BH at low voltage, (b) delayed BH (arrow) and (c) faster BH in MTJs with high volume low moment FL. (d) BH in MTJ formed from a Ta-based hybrid FL. (e, f) Ibidem for an MTJ with a dual MgO FL and with a W-based hybrid free layer. In the sole (e) case, the in-current state sometimes relaxes to P instead of AP when the current is removed. }
\label{ALL-FL-BH}
\end{figure}

Provided specific precautions are taken, the time-resolved data are adequate to demonstrate that P and P' are different states. We have therefore built the following set-up. We apply voltage steps to the MTJ from fast-rising home-designed generators. The voltage step arriving at the device had a rise time of 18 ps [Fig.~\ref{ANTHOLOGY} and \ref{ALL-FL-BH}(e)] or 1.1 ns (Fig.~\ref{ALL-FL-BH}). We heavily attenuate the output of the generators to match them and cancel their electrical reflections that would otherwise form echos at the device and generate multistep stimuli. We record the current passing through the MTJ using a high bandwidth oscilloscope. When high resolution is needed, we add amplifiers that are placed after long electrical delays that postpone triple-transit echos to out of the measurement time-window. The finite electrical bandwidth of the device, of the amplifiers and the dispersive loss of the delaying cables degrade the rise time of the transmitted current, which increases up to 120 ps in the best case. The conductance is calculated as the current/voltage ratio and therefore can only be displayed for finite voltage, i.e. after the onset of the voltage step. To avoid degraded signal-to-noise ratio and to circumvent the voltage-dependence of the conductance, we are only displaying the conductance when the voltage has reached a perfectly constant plateau, i.e. 300 ps after the pulse onset. Note that the device impedance is much greater than the characteristic impedance of its surroundings, such that changes in the device resistance do not change the applied voltage. The time-resolved conductance curves (Fig.~\ref{ANTHOLOGY}) after the first 300 ps are thus illustrative of the device dynamics at \textit{strictly constant} applied voltage. 

The device temperature deserves a comment. The current-induced Joule heating of the device is known \cite{beek_impact_2018} to lead to a temperature increase $\Delta T$ that completes within a poorly-known thermalization time $\tau_\textrm{th}$ that can be between slightly less than a nanosecond and up to at most 10 ns. $\Delta T$ depends on the Joule power density $V_\textrm{applied}^2/RA$, being typically\cite{herault_nanosecond_2009, goff_spin-wave_2016, beek_impact_2018} 2-4 K per $\textrm{mW} / \mu \textrm m^2$. With our parameters at 0.7 V (Fig.~\ref{ANTHOLOGY}), this means that the temperature rise in a steady P state would be 100 to 200K, and 50 to 100K in a steady AP state. If the thermalization time $\tau_\textrm{th}$ was shorter than the switching durations, then the magnetization switching events could be considered to result in almost coincidental changes in the device temperature, including an instantaneous device cooling at the P$\rightarrow$AP transition. However $\tau_\textrm{th}$ is not sufficiently known, such that the time-resolved conductance curves (Fig.~\ref{ANTHOLOGY}) should be considered as illustrative of both the magnetization dynamics \textit{and} the temperature dynamics. The noticeable exceptions is when the device is close to the P state because the conductance of this state is almost temperature independent (see Fig.\ref{RVloops}).

This important statement being said, let's analyse the time-resolved conductance signature of BH. The BH proceeds in a sequence of five steps. Once the applied voltage has reached its plateau, the P state (i) first stays constant and quiet during an incubation delay that lasts for a few ns, (ii) then some weak amplitude pre-dynamics occurs and lets the conductance decrease below that of P before (iii) the P$\rightarrow$AP transition proceeds at a time $t=t_\textrm{PAP}$ with a transition lasting typically $\tau_\textrm{PAP}$. (iv) No specific feature is observed in the AP state, which looks microwave quiet. Finally, (v) it takes $\tau_\textrm{APP'}$ for the device to switch to the P' state; this last step happens $t_\textrm{APP'}$ after the settling of the AP state [see Fig.~\ref{ANTHOLOGY}(a)]. The single-shot curves [Fig.~\ref{ANTHOLOGY}(a,b)] as well as the lower-noise event-averaged curves (c) for the high volume-FL samples reveal that $R_{\textrm{P'}} = R_\textrm{P} / 0.978$ at the same $V_\textrm{applied}=0.7~\textrm V$. As discussed above, this difference cannot be accounted for by Joule heating. This proves that P and P' are different states. 
%

We stress that despite the clear stochasticity of the BH process, it always proceeds through the same sequence: there is often some pre-dynamics perceivable before the P$\rightarrow$AP switching; a genuine AP state is visited in 99\% of the BH events before the system leaves it. \textcolor{black}{The remaining 1\% cases (not shown) correspond to when the AP$\rightarrow$P' transition begins soon after the P$\rightarrow$AP has started but before it is completed: it means that the AP$\rightarrow$P' transition starts to incubate as soon as a non-empty part of the FL has switched in a locally AP configuration.}
We have fitted the time-resolved conductance curves to combinations of error functions (i.e. of the form $\textrm{ERF}(t-t_{i})/\tau_{i}$ with $i\in \{ \textrm{PAP},~ \textrm{APP'} \}$) to extract the characteristic switching and transition times of the BH sequence \textcolor{black}{(ERF functions were found empirically to resemble most switching curves)}. 
The two transition times $\tau_\textrm{PAP}$ and $\tau_\textrm{APP'}$ are Gaussianly distributed around mean values of 2.1 and 2.5 ns with standard deviations being 0.58 and 0.65 ns. Comparatively, the switching times $t_\textrm{PAP}$ and $t_\textrm{APP'}$ are less stochastic with mean values of 9.9 and 8.8 ns and standard deviations of 1.9 and 2.0 ns. The distributions of $t_\textrm{PAP}$ and $t_\textrm{APP'}$ are also rather symmetric, such that the event-averaged conductance curves can be fitted also [Fig.~\ref{ANTHOLOGY}(c)] with the mean values of the switching times (9.9 and 8.8 ns). The widths of the so-fitted $\textrm{ERF}$ functions are consistent with the square sum rules of the transition times and of the standard deviations of the switching times. Note that the pre-dynamics is still perceivable even after the averaging, which reflects that despite its high degree of stochasticity, the slight conductance increase before the onset of switching is at least often (if not systematically) happening. It is important to mention that the pre-dynamics, the distributions of switching times and of transition times of the P$\rightarrow$AP and AP$\rightarrow$P' are very similar in the back-hopping-free AP$\rightarrow$P transition happening at negative voltage (not shown): they all reflect spin-torque induced switching events.

Note also the AP conductance level is not reached in the event-averaged curves only because the distribution of $t_\textrm{PAP}$ and of $t_\textrm{PAP}+t_\textrm{APP'}-\frac{1}{2} \tau_\textrm{PAP}$ have some overlap.
If the voltage step is reduced, the BH sequence gets not only less probable and slower but also surprisingly more complex. An additional state, labeled as "dyn" in Fig.~\ref{ANTHOLOGY}(d), gets often detectable as a very microwave-noisy state of conductance intermediate between AP and P. This dynamical state either appears transiently in the BH path from AP$\rightarrow$dyn$\rightarrow$P', or in a AP$\rightarrow$dyn$\rightarrow$AP failed switching attempt; or in any combination of these paths that terminates on P'.

So far we have described in detail the BH in the specific case of MTJs made with a high volume low moment FL. In fact the P$\rightarrow$AP$\rightarrow$P' main features of BH are preserved in all the MTJs that posses the fixed system depicted in Fig.~\ref{STACK}. 
The similarity is illustrated in the stack-to-stack comparison of BH curves in Fig.~\ref{ALL-FL-BH}. When the fixed system is grown on a Ta-based hybrid free layer, the BH essentially ressembles that described in detail in the high volume low moment-FL MTJ. When the fixed system is grown on a W-based hybrid FL the P' state does not seem to be very stable:  transient spikes of resistance are detected in the P' state [Fig.~\ref{ALL-FL-BH}(d)]. This apparent instability of the P' state is even more pronounced when the MTJ fixed system is grown on the dual MgO FL, and in this case, the noisy state relaxes to P instead of AP, with a measurable probability of $3\#/200\#$. We remind that for these two samples the AP-P' transition was appearing as reversible (anhysteretic) in quasi-static STT loops.

%

\section{Discussion}
\begin{table*}
	\begin{center}
		\begin{tabular}{|c|c|c||c||c|c|c}
Texture-breaker Layer  & Ta & W & theoretical & TaFeCoB & WFeCoB\\
\hline
Thickness and annealing & 3\r{A}, 300$^\circ$C & 3\r{A}, 300$^\circ$C & ferrocoupler & 8\r{A}, 300$^\circ$C & 8\r{A}, 400$^\circ$C\\
\hline

RL-SPL exchange coupling (mJ/m$^2$)  &  0.07\cite{devolder_ferromagnetic_2016}, 0.21 \cite{le_goff_effect_2014}, 0.37\cite{devolder_evolution_2016} & 0.22\cite{devolder_ferromagnetic_2016} $\ddagger$  & $J_0=0.33$ \cite{liu_control_2017} $^\dagger$ &  $J \geq1$ \cite{devolder_annealing_2017} & $J >1$ \cite{devolder_effect_2019} \\
\hline
Back-hopping is reported & yes \cite{devolder_time-resolved_2016, devolder_evolution_2016} & yes $^\textrm{this~work}$ & yes if $J < J_0$, no if $J > J_0$& no \cite{devolder_offset_2019} & no \cite{devolder_offset_2019, devolder_effect_2019} \\
\hline
		\end{tabular}
	\end{center}
    \caption{Review of the studies of STT-switching in PMA-MTJs in which the full stack details is disclosed and the back-hopping is commented on. $\dagger$: The data from ref. \onlinecite{liu_control_2017}, originally calculated for 9\r{A} of CoFeB under a current density of $10^{11}~\textrm{A/m}^2$ was rescaled to our case of 11\r{A}. $\ddagger$: this number is for a CoFeB/W/[Ni6\r{A}/Co3\r{A}]$_{\times 4}$.}
    \label{BH-tab}
\end{table*}
Before discussing them, let us summarize our experimental findings. The BH proceeds in two sequential switching events P$\rightarrow$AP and AP$\rightarrow$P' that can happen at the same voltage. 
The P' state does not exist at remanence: its generally relaxes to AP when the current is stopped. Depending on the growth conditions of the fixed system, the P' state can either be microwave quiet with a resistance slightly higher than P, or it can show a tendency for dynamic instability.

\subsection{Identity of the P' state of the back-hopping} 
Two scenarios were proposed in the past to explain the BH phenomenon. The scenario proposed in ref. \onlinecite{skowronski_backhopping_2013, frankowski_backhopping_2015} was based on a bias-dependent sign reversal of the STT acting on the FL:  at low positive bias the STT would induce the desired P$\rightarrow$AP transition, while at large bias the sign of STT would reverse and induce the undesired AP$\rightarrow$P back transition with the possible assistance of some field-like torque. There was no P' state in that proposed scenario. 
Our results show that P$\rightarrow$AP and AP$\rightarrow$P' happen sequentially at the same voltage bias. The spin polarization transferred from a \textit{fixed} SPL to the FL can obviously not be first positive then negative a few ns after at the \textit{same} voltage; we conclude that the BH observed in our study is not the switching back of the free layer that would result from an hypothetical strongly anomalous bias dependence of the STT. 

We therefore turn to the second proposed scenario\cite{devolder_time-resolved_2016, abert_back-hopping_2018, hu_reliable_2019, safranski_interface_2019}, which suggested that some "sub-system" within the fixed layers system is destabilized by STT once the FL has switched.
This sub-system affects the MTJ resistance and must thus include a switching of at least the SPL. The switching sub-system can thus either be: \\(i) the whole fixed system, i.e. a rigidly bound \{SPL + RL + HL\} ensemble, \\(ii) or a rigidly bound \{SPL + RL\} ensemble or, \\(iii) the sole SPL.\\ The first option can be straightforwardly ruled out since a fully switched fixed system would not re-switch back upon current reduction in the R(V) loops. Such a state would also be immediately evidenced as a subsequent change of the R(H) minor loop, which is not observed. The second and third options are both conceivable since the antiferromagnetic (respectively ferromagnetic) coupling though Ru (resp. W) could cause the \{SPL+RL\} ensemble (resp. the sole SPL) to switch back upon current reduction. We believe that the last option is the most likely: indeed the weak link within the fixed system is the texture-breaker ferro-coupling layer (W 3\r{A} in our case):  the interlayer exchange coupling though W ($J_\textrm W=0.22~\textrm{mJ/m}^2$, see Table~\ref{BH-tab}) is much weaker than through an 8.5\r{A} Ru layer annealed at 300$^\circ$C, which typically \cite{devolder_evolution_2016} amounts to $J_\textrm{Ru}= -1.3~\textrm{mJ/m}^2$. 

We thus conjecture that the P' state corresponds to a modified AP state in which the magnetization of the sole SPL is reversed [see the arrows in Fig.~\ref{STACK}]. We believe that the resistance of the P' state is higher than that of the P state simply because the depth of the zones with magnetization parallel to the FL within the fixed system are different: only 11\r{A} for P' versus 11+3+6 \r{A} for P. It is indeed known \cite{cuchet_influence_2014} that 17 \r{A} of FeCoB is needed to asymptotically reach the full TMR potential of a reference layer. 

\subsection{Role of the different magnetic properties in the back-hopping process}
Our conjecture that the P' state arises from the switching of the sole spin-polarizing CoFeB 11 \r{A} SPL layer is also supported by a careful analysis of the reports on BH along the historical evolution of STT-MRAM in the recent years.
Our point is the following: insufficient stability within the fixed system was identified as a major stack deficiency in the years 2013-2016, which triggered searches for higher anisotropy and higher coupling within the fixed system. Higher anisotropy was achieved by passing from Co/Pd multilayers \cite{worledge_spin_2011, gajek_spin_2012} to Co/Ni \cite{tomczak_thin_2016} and finally ubiquitously to Co/Pt. Higher coupling was achieved by changing the antiferrocoupler from Ru 8.5\r{A} to Ru 4\r{A} \cite{couet_impact_2017} and finally Ir 5.2\r{A} \cite{devolder_offset_2019}, but also by thinning the RL \cite{tomczak_thin_2016} to increase its pinning field. However these material optimizations stabilized the RL \textit{but not} the SPL and were not reported to solve the BH problem. 

Looking back at history (Table~\ref{BH-tab}), it appears that solving the BH problem was obtained by progresses in the texture-transition layer that was changed successively from ultrathin (3\r{A}) Ta to W, then to alloys of refractory metals and magnetic metals \cite{gottwald_paramagnetic_2013}, e.g.  TaFeCoB and WFeCoB in which the almost ferromagnetic character enabled a much higher exchange coupling, as well as \cite{kim_ultrathin_2015, liu_high_2016, couet_impact_2017} an increase of anisotropy in the case of W-based alloys. Both TaFeCoB and WFeCoB spacers appeared to suppress BH, in addition to providing a larger process window (6 to 8 \r{A} of thickness instead of 3). 

Note that in a different context, E. Liu et al. \cite{liu_control_2017} performed a micromagnetic study to determine how much ferromagnetic interlayer coupling would be needed to maintain the parallel alignement of two PMA ferromagnets when the top one is subjected to STT. E. Liu et al.  concluded that for a current density of $10^{11}~\textrm{A/m}^2$, a coupling larger than $0.27~\textrm{mJ/m}^2$ was needed to lock a top-positioned CoFeB 9 \r{A} layer parallel to the other one. This value, when renormalized to the thickness of our SPL to achieve the same exchange field, nicely fits in between the situations in which BH was or was not experimentally observed (see Table~\ref{BH-tab}.) This comforts us in the conjecture that BH is due to the undesired switching of the sole SPL.

\subsection{Link between dynamical back-hopping and write error rate} 
Let us comment on the link between back-hopping and write error rate. The degree of stability of the P' state determines whether an MTJ stays in this state when the voltage is let applied. If P' is not sufficiently stable, there can be a subsequent evolution after the P$\rightarrow$AP$\rightarrow$P' steps: the STT could destabilize the FL of the P' state and switch the MTJ to an hypothetical AP' state [see Fig.~\ref{STACK}(a)]. Then the STT acting on the SPL would easily switch the SPL so that the MTJ would recover the true P state, forming a cycle that can start anew and continue indefinitely.  

In this configuration, write errors are expected since the state obtained after current removal is determined by where the system is positioned in the perpetual sequence [P$\rightarrow$AP$\rightarrow$P'$\rightarrow$AP'$\rightarrow \textrm P]_{\times \infty}$ when the current is switched off; (This was also concluded in ref.~\onlinecite{hu_reliable_2019} from other arguments). We believe that the P'$\rightarrow$AP' is not likely to happen in our samples simply because the \{SPL+RL\} ensemble is a poor spin-polarizer when in the P' state, so that it does not easily trigger the switching of the FL to the hypothetical AP' state. The apparent noise in the P' states of the dual MgO FL and the W-spacer based FL [Fig.\ref{ALL-FL-BH}(e,f)] might be reminiscent of most often failing P'$\rightarrow$AP' switching attempts. Indeed the appearance of a strong apparent noise in the in-current state correlates with the onset of a non-vanishing write error rate (1.5\%) measured at the highest voltage on the dual MgO FL samples.

As a side remark, we would like to mention that a \textit{mirror} back-hopping scenario can be envisioned for the AP$\rightarrow$P transition, by following the sequence [AP$\rightarrow$P$\rightarrow$AP'$\rightarrow$P'$\rightarrow \textrm {AP}]_{\times \infty}$. We do not observe this scenario with the present samples, but this can be conjectured from the shape of the R(V) loops when applying a field assistance to destabilize deliberately the SPL \cite{hu_reliable_2019, devolder_size_2016}. 

\subsection{Material options of the mitigation of the back-hopping}
This understanding of the BH phenomenon can be used to define material improvements that would minimize high bias write error rates. Although BH stems from a failure of the SPL to keep a fixed magnetization, the improvements can involve both the fixed system and the free layer system. 

Within the fixed system, three directions can be followed: the strengthening of the SPL by a strong exchange coupling with the RL, the strengthening of the SPL by a maximization of its anisotropy, and the minimization of the SPL susceptibility to STT by an increase of its Gilbert damping. These three points argue for the insertion of a WCoFeB-based texture transition layers at the SPL-RL interface, which seems to effectively mitigate the back-hopping issue \cite{devolder_offset_2019}. 

Within the FL, the rational for material optimization could be the following. The desired transition (i.e. P$\rightarrow$AP at $V_\textrm{app}>0$) is induced by STT acting on the FL which  --provided separability applies (see ref.~\onlinecite{slonczewski_currents_2005}, section IV-- results from the spin polarization of the SPL.  Conversely, the undesired transition (i.e. AP$\rightarrow$P' at $V_\textrm{app}>0$) responsible for BH is induced by STT acting on the SPL resulting from the spin polarization of the FL (provided once again that separability applies). To mitigate back-hopping, one could thus think of designing the FL to reduce the spin-polarization of its outgoing tunneling electrons, for instance by tuning the spacer layer inserted at the midst of the free layer.

\section{Conclusion}
We have studied of the phenomenon of back-hopping in the spin-torque switching in perpendicularly magnetized tunnel junctions. Our analysis relies on single-shot time-resolved conductance measurements of the voltage-pulse-induced back-hopping in various tunnel junctions sharing a common fixed system. The back-hopping is found to proceed by two sequential switching events P$\rightarrow$AP and AP$\rightarrow$P' that occur at the same voltage and lead to a final state P' of conductance close to --but distinct from-- that of the conventional parallel state. This rules out back-hopping explanations based on the switching back of the free layer as a result of some anomalous voltage dependence of the spin torques acting on the free layer. 

The P' state involves in fact a switching of the sole spin-polarizing part of the fixed layers. The back-hopping occurs only when the spin-polarizing layer is too weakly coupled to the rest of the fixed system. We conjecture that the back hopping can either stop in the P' state or undergo further evolutions with a cyclic return to the P state, depending on the ability of the reference layer to supply enough spin-torque when in the P' state. Our results shed light on the mitigation strategies of back-hopping that were implemented empirically in spin-transfer-torque magnetic random access memories. 

Acknowledgement: this work was supported in part by the IMEC’s Industrial Affiliation Program on STT-MRAM device, and in part by the imec IIAP core CMOS and the Beyond CMOS program of Intel Corporation. T.D. and P. B. thank Jonathan Z. Sun for constructive discussions on the BH phenomenon.


\begin{thebibliography}{33}%
\makeatletter
\providecommand \@ifxundefined [1]{%
 \@ifx{#1\undefined}
}%
\providecommand \@ifnum [1]{%
 \ifnum #1\expandafter \@firstoftwo
 \else \expandafter \@secondoftwo
 \fi
}%
\providecommand \@ifx [1]{%
 \ifx #1\expandafter \@firstoftwo
 \else \expandafter \@secondoftwo
 \fi
}%
\providecommand \natexlab [1]{#1}%
\providecommand \enquote  [1]{``#1''}%
\providecommand \bibnamefont  [1]{#1}%
\providecommand \bibfnamefont [1]{#1}%
\providecommand \citenamefont [1]{#1}%
\providecommand \href@noop [0]{\@secondoftwo}%
\providecommand \href [0]{\begingroup \@sanitize@url \@href}%
\providecommand \@href[1]{\@@startlink{#1}\@@href}%
\providecommand \@@href[1]{\endgroup#1\@@endlink}%
\providecommand \@sanitize@url [0]{\catcode `\\12\catcode `\$12\catcode
  `\&12\catcode `\#12\catcode `\^12\catcode `\_12\catcode `\%12\relax}%
\providecommand \@@startlink[1]{}%
\providecommand \@@endlink[0]{}%
\providecommand \url  [0]{\begingroup\@sanitize@url \@url }%
\providecommand \@url [1]{\endgroup\@href {#1}{\urlprefix }}%
\providecommand \urlprefix  [0]{URL }%
\providecommand \Eprint [0]{\href }%
\providecommand \doibase [0]{http://dx.doi.org/}%
\providecommand \selectlanguage [0]{\@gobble}%
\providecommand \bibinfo  [0]{\@secondoftwo}%
\providecommand \bibfield  [0]{\@secondoftwo}%
\providecommand \translation [1]{[#1]}%
\providecommand \BibitemOpen [0]{}%
\providecommand \bibitemStop [0]{}%
\providecommand \bibitemNoStop [0]{.\EOS\space}%
\providecommand \EOS [0]{\spacefactor3000\relax}%
\providecommand \BibitemShut  [1]{\csname bibitem#1\endcsname}%
\let\auto@bib@innerbib\@empty
\bibitem [{\citenamefont {Khvalkovskiy}\ \emph {et~al.}(2013)\citenamefont
  {Khvalkovskiy}, \citenamefont {Apalkov}, \citenamefont {Watts}, \citenamefont
  {Chepulskii}, \citenamefont {Beach}, \citenamefont {Ong}, \citenamefont
  {Tang}, \citenamefont {Driskill-Smith}, \citenamefont {Butler}, \citenamefont
  {Visscher}, \citenamefont {Lottis}, \citenamefont {Chen}, \citenamefont
  {Nikitin},\ and\ \citenamefont {Krounbi}}]{khvalkovskiy_basic_2013}%
  \BibitemOpen
  \bibfield  {author} {\bibinfo {author} {\bibfnamefont {A.~V.}\ \bibnamefont
  {Khvalkovskiy}}, \bibinfo {author} {\bibfnamefont {D.}~\bibnamefont
  {Apalkov}}, \bibinfo {author} {\bibfnamefont {S.}~\bibnamefont {Watts}},
  \bibinfo {author} {\bibfnamefont {R.}~\bibnamefont {Chepulskii}}, \bibinfo
  {author} {\bibfnamefont {R.~S.}\ \bibnamefont {Beach}}, \bibinfo {author}
  {\bibfnamefont {A.}~\bibnamefont {Ong}}, \bibinfo {author} {\bibfnamefont
  {X.}~\bibnamefont {Tang}}, \bibinfo {author} {\bibfnamefont {A.}~\bibnamefont
  {Driskill-Smith}}, \bibinfo {author} {\bibfnamefont {W.~H.}\ \bibnamefont
  {Butler}}, \bibinfo {author} {\bibfnamefont {P.~B.}\ \bibnamefont
  {Visscher}}, \bibinfo {author} {\bibfnamefont {D.}~\bibnamefont {Lottis}},
  \bibinfo {author} {\bibfnamefont {E.}~\bibnamefont {Chen}}, \bibinfo {author}
  {\bibfnamefont {V.}~\bibnamefont {Nikitin}}, \ and\ \bibinfo {author}
  {\bibfnamefont {M.}~\bibnamefont {Krounbi}},\ }\href {\doibase
  10.1088/0022-3727/46/7/074001} {\bibfield  {journal} {\bibinfo  {journal}
  {Journal of Physics D: Applied Physics}\ }\textbf {\bibinfo {volume} {46}},\
  \bibinfo {pages} {074001} (\bibinfo {year} {2013})}\BibitemShut {NoStop}%
\bibitem [{\citenamefont {Hu}\ \emph {et~al.}(2019)\citenamefont {Hu},
  \citenamefont {Nowak}, \citenamefont {Gottwald}, \citenamefont {Sun},
  \citenamefont {Houssameddine}, \citenamefont {Bak}, \citenamefont {Brown},
  \citenamefont {Hashemi}, \citenamefont {He}, \citenamefont {Kim},
  \citenamefont {Kothandaraman}, \citenamefont {Lauer}, \citenamefont {Lee},
  \citenamefont {Suwannasiri}, \citenamefont {Trouilloud},\ and\ \citenamefont
  {Worledge}}]{hu_reliable_2019}%
  \BibitemOpen
  \bibfield  {author} {\bibinfo {author} {\bibfnamefont {G.}~\bibnamefont
  {Hu}}, \bibinfo {author} {\bibfnamefont {J.~J.}\ \bibnamefont {Nowak}},
  \bibinfo {author} {\bibfnamefont {M.~G.}\ \bibnamefont {Gottwald}}, \bibinfo
  {author} {\bibfnamefont {J.~Z.}\ \bibnamefont {Sun}}, \bibinfo {author}
  {\bibfnamefont {D.}~\bibnamefont {Houssameddine}}, \bibinfo {author}
  {\bibfnamefont {J.}~\bibnamefont {Bak}}, \bibinfo {author} {\bibfnamefont
  {S.~L.}\ \bibnamefont {Brown}}, \bibinfo {author} {\bibfnamefont
  {P.}~\bibnamefont {Hashemi}}, \bibinfo {author} {\bibfnamefont
  {Q.}~\bibnamefont {He}}, \bibinfo {author} {\bibfnamefont {J.}~\bibnamefont
  {Kim}}, \bibinfo {author} {\bibfnamefont {C.}~\bibnamefont {Kothandaraman}},
  \bibinfo {author} {\bibfnamefont {G.}~\bibnamefont {Lauer}}, \bibinfo
  {author} {\bibfnamefont {H.~K.}\ \bibnamefont {Lee}}, \bibinfo {author}
  {\bibfnamefont {T.}~\bibnamefont {Suwannasiri}}, \bibinfo {author}
  {\bibfnamefont {P.~L.}\ \bibnamefont {Trouilloud}}, \ and\ \bibinfo {author}
  {\bibfnamefont {D.~C.}\ \bibnamefont {Worledge}},\ }\href {\doibase
  10.1109/LMAG.2019.2928243} {\bibfield  {journal} {\bibinfo  {journal} {IEEE
  Magnetics Letters}\ }\textbf {\bibinfo {volume} {10}},\ \bibinfo {pages} {1}
  (\bibinfo {year} {2019})},\ \bibinfo {note} {conference Name: IEEE Magnetics
  Letters}\BibitemShut {NoStop}%
\bibitem [{\citenamefont {Min}\ \emph {et~al.}(2009)\citenamefont {Min},
  \citenamefont {Sun}, \citenamefont {Beach}, \citenamefont {Tang},\ and\
  \citenamefont {Wang}}]{min_back-hopping_2009}%
  \BibitemOpen
  \bibfield  {author} {\bibinfo {author} {\bibfnamefont {T.}~\bibnamefont
  {Min}}, \bibinfo {author} {\bibfnamefont {J.~Z.}\ \bibnamefont {Sun}},
  \bibinfo {author} {\bibfnamefont {R.}~\bibnamefont {Beach}}, \bibinfo
  {author} {\bibfnamefont {D.}~\bibnamefont {Tang}}, \ and\ \bibinfo {author}
  {\bibfnamefont {P.}~\bibnamefont {Wang}},\ }\href {\doibase
  10.1063/1.3063672} {\bibfield  {journal} {\bibinfo  {journal} {Journal of
  Applied Physics}\ }\textbf {\bibinfo {volume} {105}},\ \bibinfo {pages}
  {07D126} (\bibinfo {year} {2009})}\BibitemShut {NoStop}%
\bibitem [{\citenamefont {Sun}\ \emph {et~al.}(2009)\citenamefont {Sun},
  \citenamefont {Gaidis}, \citenamefont {Hu}, \citenamefont {O’Sullivan},
  \citenamefont {Brown}, \citenamefont {Nowak}, \citenamefont {Trouilloud},\
  and\ \citenamefont {Worledge}}]{sun_high-bias_2009}%
  \BibitemOpen
  \bibfield  {author} {\bibinfo {author} {\bibfnamefont {J.~Z.}\ \bibnamefont
  {Sun}}, \bibinfo {author} {\bibfnamefont {M.~C.}\ \bibnamefont {Gaidis}},
  \bibinfo {author} {\bibfnamefont {G.}~\bibnamefont {Hu}}, \bibinfo {author}
  {\bibfnamefont {E.~J.}\ \bibnamefont {O’Sullivan}}, \bibinfo {author}
  {\bibfnamefont {S.~L.}\ \bibnamefont {Brown}}, \bibinfo {author}
  {\bibfnamefont {J.~J.}\ \bibnamefont {Nowak}}, \bibinfo {author}
  {\bibfnamefont {P.~L.}\ \bibnamefont {Trouilloud}}, \ and\ \bibinfo {author}
  {\bibfnamefont {D.~C.}\ \bibnamefont {Worledge}},\ }\href {\doibase
  10.1063/1.3058614} {\bibfield  {journal} {\bibinfo  {journal} {Journal of
  Applied Physics}\ }\textbf {\bibinfo {volume} {105}},\ \bibinfo {pages}
  {07D109} (\bibinfo {year} {2009})},\ \bibinfo {note} {publisher: American
  Institute of Physics}\BibitemShut {NoStop}%
\bibitem [{\citenamefont {Skowroński}\ \emph {et~al.}(2013)\citenamefont
  {Skowroński}, \citenamefont {Ogrodnik}, \citenamefont {Wrona}, \citenamefont
  {Stobiecki}, \citenamefont {Świrkowicz}, \citenamefont {Barnaś},
  \citenamefont {Reiss},\ and\ \citenamefont
  {Dijken}}]{skowronski_backhopping_2013}%
  \BibitemOpen
  \bibfield  {author} {\bibinfo {author} {\bibfnamefont {W.}~\bibnamefont
  {Skowroński}}, \bibinfo {author} {\bibfnamefont {P.}~\bibnamefont
  {Ogrodnik}}, \bibinfo {author} {\bibfnamefont {J.}~\bibnamefont {Wrona}},
  \bibinfo {author} {\bibfnamefont {T.}~\bibnamefont {Stobiecki}}, \bibinfo
  {author} {\bibfnamefont {R.}~\bibnamefont {Świrkowicz}}, \bibinfo {author}
  {\bibfnamefont {J.}~\bibnamefont {Barnaś}}, \bibinfo {author} {\bibfnamefont
  {G.}~\bibnamefont {Reiss}}, \ and\ \bibinfo {author} {\bibfnamefont {S.~v.}\
  \bibnamefont {Dijken}},\ }\href {\doibase 10.1063/1.4843635} {\bibfield
  {journal} {\bibinfo  {journal} {Journal of Applied Physics}\ }\textbf
  {\bibinfo {volume} {114}},\ \bibinfo {pages} {233905} (\bibinfo {year}
  {2013})},\ \bibinfo {note} {publisher: American Institute of
  Physics}\BibitemShut {NoStop}%
\bibitem [{\citenamefont {Frankowski}\ \emph {et~al.}(2015)\citenamefont
  {Frankowski}, \citenamefont {Skowroński}, \citenamefont {Czapkiewicz},\ and\
  \citenamefont {Stobiecki}}]{frankowski_backhopping_2015}%
  \BibitemOpen
  \bibfield  {author} {\bibinfo {author} {\bibfnamefont {M.}~\bibnamefont
  {Frankowski}}, \bibinfo {author} {\bibfnamefont {W.}~\bibnamefont
  {Skowroński}}, \bibinfo {author} {\bibfnamefont {M.}~\bibnamefont
  {Czapkiewicz}}, \ and\ \bibinfo {author} {\bibfnamefont {T.}~\bibnamefont
  {Stobiecki}},\ }\href {\doibase 10.1016/j.jmmm.2014.08.056} {\bibfield
  {journal} {\bibinfo  {journal} {Journal of Magnetism and Magnetic Materials}\
  }\textbf {\bibinfo {volume} {374}},\ \bibinfo {pages} {451} (\bibinfo {year}
  {2015})}\BibitemShut {NoStop}%
\bibitem [{\citenamefont {Theodonis}\ \emph {et~al.}(2006)\citenamefont
  {Theodonis}, \citenamefont {Kioussis}, \citenamefont {Kalitsov},
  \citenamefont {Chshiev},\ and\ \citenamefont
  {Butler}}]{theodonis_anomalous_2006}%
  \BibitemOpen
  \bibfield  {author} {\bibinfo {author} {\bibfnamefont {I.}~\bibnamefont
  {Theodonis}}, \bibinfo {author} {\bibfnamefont {N.}~\bibnamefont {Kioussis}},
  \bibinfo {author} {\bibfnamefont {A.}~\bibnamefont {Kalitsov}}, \bibinfo
  {author} {\bibfnamefont {M.}~\bibnamefont {Chshiev}}, \ and\ \bibinfo
  {author} {\bibfnamefont {W.~H.}\ \bibnamefont {Butler}},\ }\href {\doibase
  10.1103/PhysRevLett.97.237205} {\bibfield  {journal} {\bibinfo  {journal}
  {Physical Review Letters}\ }\textbf {\bibinfo {volume} {97}},\ \bibinfo
  {pages} {237205} (\bibinfo {year} {2006})}\BibitemShut {NoStop}%
\bibitem [{\citenamefont {Devolder}\ \emph
  {et~al.}(2016{\natexlab{a}})\citenamefont {Devolder}, \citenamefont {Kim},
  \citenamefont {Garcia-Sanchez}, \citenamefont {Swerts}, \citenamefont {Kim},
  \citenamefont {Couet}, \citenamefont {Kar},\ and\ \citenamefont
  {Furnemont}}]{devolder_time-resolved_2016}%
  \BibitemOpen
  \bibfield  {author} {\bibinfo {author} {\bibfnamefont {T.}~\bibnamefont
  {Devolder}}, \bibinfo {author} {\bibfnamefont {J.-V.}\ \bibnamefont {Kim}},
  \bibinfo {author} {\bibfnamefont {F.}~\bibnamefont {Garcia-Sanchez}},
  \bibinfo {author} {\bibfnamefont {J.}~\bibnamefont {Swerts}}, \bibinfo
  {author} {\bibfnamefont {W.}~\bibnamefont {Kim}}, \bibinfo {author}
  {\bibfnamefont {S.}~\bibnamefont {Couet}}, \bibinfo {author} {\bibfnamefont
  {G.}~\bibnamefont {Kar}}, \ and\ \bibinfo {author} {\bibfnamefont
  {A.}~\bibnamefont {Furnemont}},\ }\href {\doibase 10.1103/PhysRevB.93.024420}
  {\bibfield  {journal} {\bibinfo  {journal} {Physical Review B}\ }\textbf
  {\bibinfo {volume} {93}},\ \bibinfo {pages} {024420} (\bibinfo {year}
  {2016}{\natexlab{a}})}\BibitemShut {NoStop}%
\bibitem [{\citenamefont {Safranski}\ and\ \citenamefont
  {Sun}(2019)}]{safranski_interface_2019}%
  \BibitemOpen
  \bibfield  {author} {\bibinfo {author} {\bibfnamefont {C.}~\bibnamefont
  {Safranski}}\ and\ \bibinfo {author} {\bibfnamefont {J.~Z.}\ \bibnamefont
  {Sun}},\ }\href {\doibase 10.1103/PhysRevB.100.014435} {\bibfield  {journal}
  {\bibinfo  {journal} {Physical Review B}\ }\textbf {\bibinfo {volume}
  {100}},\ \bibinfo {pages} {014435} (\bibinfo {year} {2019})},\ \bibinfo
  {note} {arXiv: 1905.13253}\BibitemShut {NoStop}%
\bibitem [{\citenamefont {Abert}\ \emph {et~al.}(2018)\citenamefont {Abert},
  \citenamefont {Sepehri-Amin}, \citenamefont {Bruckner}, \citenamefont
  {Vogler}, \citenamefont {Hayashi},\ and\ \citenamefont
  {Suess}}]{abert_back-hopping_2018}%
  \BibitemOpen
  \bibfield  {author} {\bibinfo {author} {\bibfnamefont {C.}~\bibnamefont
  {Abert}}, \bibinfo {author} {\bibfnamefont {H.}~\bibnamefont {Sepehri-Amin}},
  \bibinfo {author} {\bibfnamefont {F.}~\bibnamefont {Bruckner}}, \bibinfo
  {author} {\bibfnamefont {C.}~\bibnamefont {Vogler}}, \bibinfo {author}
  {\bibfnamefont {M.}~\bibnamefont {Hayashi}}, \ and\ \bibinfo {author}
  {\bibfnamefont {D.}~\bibnamefont {Suess}},\ }\href {\doibase
  10.1103/PhysRevApplied.9.054010} {\bibfield  {journal} {\bibinfo  {journal}
  {Physical Review Applied}\ }\textbf {\bibinfo {volume} {9}},\ \bibinfo
  {pages} {054010} (\bibinfo {year} {2018})},\ \bibinfo {note} {publisher:
  American Physical Society}\BibitemShut {NoStop}%
\bibitem [{\citenamefont {Thomas}\ \emph {et~al.}(2017)\citenamefont {Thomas},
  \citenamefont {Benzaouia}, \citenamefont {Serrano-Guisan}, \citenamefont
  {Jan}, \citenamefont {Le}, \citenamefont {Lee}, \citenamefont {Liu},
  \citenamefont {Zhu}, \citenamefont {Iwata-Harms}, \citenamefont {Tong},
  \citenamefont {Yang}, \citenamefont {Sundar}, \citenamefont {Patel},
  \citenamefont {Haq}, \citenamefont {Shen}, \citenamefont {He}, \citenamefont
  {Lam}, \citenamefont {Teng}, \citenamefont {Liu}, \citenamefont {Wang},
  \citenamefont {Zhong}, \citenamefont {Torng},\ and\ \citenamefont
  {Wang}}]{thomas_spin_2017}%
  \BibitemOpen
  \bibfield  {author} {\bibinfo {author} {\bibfnamefont {L.}~\bibnamefont
  {Thomas}}, \bibinfo {author} {\bibfnamefont {M.}~\bibnamefont {Benzaouia}},
  \bibinfo {author} {\bibfnamefont {S.}~\bibnamefont {Serrano-Guisan}},
  \bibinfo {author} {\bibfnamefont {G.}~\bibnamefont {Jan}}, \bibinfo {author}
  {\bibfnamefont {S.}~\bibnamefont {Le}}, \bibinfo {author} {\bibfnamefont
  {Y.}~\bibnamefont {Lee}}, \bibinfo {author} {\bibfnamefont {H.}~\bibnamefont
  {Liu}}, \bibinfo {author} {\bibfnamefont {J.}~\bibnamefont {Zhu}}, \bibinfo
  {author} {\bibfnamefont {J.}~\bibnamefont {Iwata-Harms}}, \bibinfo {author}
  {\bibfnamefont {R.}~\bibnamefont {Tong}}, \bibinfo {author} {\bibfnamefont
  {Y.}~\bibnamefont {Yang}}, \bibinfo {author} {\bibfnamefont {V.}~\bibnamefont
  {Sundar}}, \bibinfo {author} {\bibfnamefont {S.}~\bibnamefont {Patel}},
  \bibinfo {author} {\bibfnamefont {J.}~\bibnamefont {Haq}}, \bibinfo {author}
  {\bibfnamefont {D.}~\bibnamefont {Shen}}, \bibinfo {author} {\bibfnamefont
  {R.}~\bibnamefont {He}}, \bibinfo {author} {\bibfnamefont {V.}~\bibnamefont
  {Lam}}, \bibinfo {author} {\bibfnamefont {J.}~\bibnamefont {Teng}}, \bibinfo
  {author} {\bibfnamefont {P.}~\bibnamefont {Liu}}, \bibinfo {author}
  {\bibfnamefont {A.}~\bibnamefont {Wang}}, \bibinfo {author} {\bibfnamefont
  {T.}~\bibnamefont {Zhong}}, \bibinfo {author} {\bibfnamefont
  {T.}~\bibnamefont {Torng}}, \ and\ \bibinfo {author} {\bibfnamefont
  {P.}~\bibnamefont {Wang}},\ }in\ \href {\doibase 10.1109/INTMAG.2017.8007621}
  {\emph {\bibinfo {booktitle} {2017 {IEEE} {International} {Magnetics}
  {Conference} ({INTERMAG})}}}\ (\bibinfo {year} {2017})\ pp.\ \bibinfo {pages}
  {1--1},\ \bibinfo {note} {iSSN: 2150-4601}\BibitemShut {NoStop}%
\bibitem [{\citenamefont {Devolder}(2016)}]{devolder_ferromagnetic_2016}%
  \BibitemOpen
  \bibfield  {author} {\bibinfo {author} {\bibfnamefont {T.}~\bibnamefont
  {Devolder}},\ }\href {\doibase 10.1063/1.4947227} {\bibfield  {journal}
  {\bibinfo  {journal} {Journal of Applied Physics}\ }\textbf {\bibinfo
  {volume} {119}},\ \bibinfo {pages} {153905} (\bibinfo {year}
  {2016})}\BibitemShut {NoStop}%
\bibitem [{\citenamefont {Hahn}\ \emph {et~al.}(2016)\citenamefont {Hahn},
  \citenamefont {Wolf}, \citenamefont {Kardasz}, \citenamefont {Watts},
  \citenamefont {Pinarbasi},\ and\ \citenamefont
  {Kent}}]{hahn_time-resolved_2016}%
  \BibitemOpen
  \bibfield  {author} {\bibinfo {author} {\bibfnamefont {C.}~\bibnamefont
  {Hahn}}, \bibinfo {author} {\bibfnamefont {G.}~\bibnamefont {Wolf}}, \bibinfo
  {author} {\bibfnamefont {B.}~\bibnamefont {Kardasz}}, \bibinfo {author}
  {\bibfnamefont {S.}~\bibnamefont {Watts}}, \bibinfo {author} {\bibfnamefont
  {M.}~\bibnamefont {Pinarbasi}}, \ and\ \bibinfo {author} {\bibfnamefont
  {A.~D.}\ \bibnamefont {Kent}},\ }\href {\doibase 10.1103/PhysRevB.94.214432}
  {\bibfield  {journal} {\bibinfo  {journal} {Physical Review B}\ }\textbf
  {\bibinfo {volume} {94}},\ \bibinfo {pages} {214432} (\bibinfo {year}
  {2016})}\BibitemShut {NoStop}%
\bibitem [{\citenamefont {Devolder}\ \emph {et~al.}(2017)\citenamefont
  {Devolder}, \citenamefont {Couet}, \citenamefont {Swerts}, \citenamefont
  {Liu}, \citenamefont {Lin}, \citenamefont {Mertens}, \citenamefont
  {Furnemont},\ and\ \citenamefont {Kar}}]{devolder_annealing_2017}%
  \BibitemOpen
  \bibfield  {author} {\bibinfo {author} {\bibfnamefont {T.}~\bibnamefont
  {Devolder}}, \bibinfo {author} {\bibfnamefont {S.}~\bibnamefont {Couet}},
  \bibinfo {author} {\bibfnamefont {J.}~\bibnamefont {Swerts}}, \bibinfo
  {author} {\bibfnamefont {E.}~\bibnamefont {Liu}}, \bibinfo {author}
  {\bibfnamefont {T.}~\bibnamefont {Lin}}, \bibinfo {author} {\bibfnamefont
  {S.}~\bibnamefont {Mertens}}, \bibinfo {author} {\bibfnamefont
  {A.}~\bibnamefont {Furnemont}}, \ and\ \bibinfo {author} {\bibfnamefont
  {G.}~\bibnamefont {Kar}},\ }\href {\doibase 10.1063/1.4978633} {\bibfield
  {journal} {\bibinfo  {journal} {Journal of Applied Physics}\ }\textbf
  {\bibinfo {volume} {121}},\ \bibinfo {pages} {113904} (\bibinfo {year}
  {2017})}\BibitemShut {NoStop}%
\bibitem [{\citenamefont {Bultynck}\ \emph {et~al.}(2018)\citenamefont
  {Bultynck}, \citenamefont {Manfrini}, \citenamefont {Vaysset}, \citenamefont
  {Swerts}, \citenamefont {Wilson}, \citenamefont {Sorée}, \citenamefont
  {Heyns}, \citenamefont {Mocuta}, \citenamefont {Radu},\ and\ \citenamefont
  {Devolder}}]{bultynck_instant-spin_2018}%
  \BibitemOpen
  \bibfield  {author} {\bibinfo {author} {\bibfnamefont {O.}~\bibnamefont
  {Bultynck}}, \bibinfo {author} {\bibfnamefont {M.}~\bibnamefont {Manfrini}},
  \bibinfo {author} {\bibfnamefont {A.}~\bibnamefont {Vaysset}}, \bibinfo
  {author} {\bibfnamefont {J.}~\bibnamefont {Swerts}}, \bibinfo {author}
  {\bibfnamefont {C.~J.}\ \bibnamefont {Wilson}}, \bibinfo {author}
  {\bibfnamefont {B.}~\bibnamefont {Sorée}}, \bibinfo {author} {\bibfnamefont
  {M.}~\bibnamefont {Heyns}}, \bibinfo {author} {\bibfnamefont
  {D.}~\bibnamefont {Mocuta}}, \bibinfo {author} {\bibfnamefont {I.~P.}\
  \bibnamefont {Radu}}, \ and\ \bibinfo {author} {\bibfnamefont
  {T.}~\bibnamefont {Devolder}},\ }\href {\doibase
  10.1103/PhysRevApplied.10.054028} {\bibfield  {journal} {\bibinfo  {journal}
  {Physical Review Applied}\ }\textbf {\bibinfo {volume} {10}},\ \bibinfo
  {pages} {054028} (\bibinfo {year} {2018})},\ \bibinfo {note} {publisher:
  American Physical Society}\BibitemShut {NoStop}%
\bibitem [{\citenamefont {Devolder}\ \emph
  {et~al.}(2019{\natexlab{a}})\citenamefont {Devolder}, \citenamefont {Couet},
  \citenamefont {Swerts}, \citenamefont {Mertens}, \citenamefont {Rao},\ and\
  \citenamefont {Kar}}]{devolder_effect_2019}%
  \BibitemOpen
  \bibfield  {author} {\bibinfo {author} {\bibfnamefont {T.}~\bibnamefont
  {Devolder}}, \bibinfo {author} {\bibfnamefont {S.}~\bibnamefont {Couet}},
  \bibinfo {author} {\bibfnamefont {J.}~\bibnamefont {Swerts}}, \bibinfo
  {author} {\bibfnamefont {S.}~\bibnamefont {Mertens}}, \bibinfo {author}
  {\bibfnamefont {S.}~\bibnamefont {Rao}}, \ and\ \bibinfo {author}
  {\bibfnamefont {G.~S.}\ \bibnamefont {Kar}},\ }\href {\doibase
  10.1109/LMAG.2019.2940572} {\bibfield  {journal} {\bibinfo  {journal} {IEEE
  Magnetics Letters}\ }\textbf {\bibinfo {volume} {10}},\ \bibinfo {pages} {1}
  (\bibinfo {year} {2019}{\natexlab{a}})},\ \bibinfo {note} {conference Name:
  IEEE Magnetics Letters}\BibitemShut {NoStop}%
\bibitem [{\citenamefont {Beek}\ \emph {et~al.}(2018)\citenamefont {Beek},
  \citenamefont {O'Sullivan}, \citenamefont {Roussel}, \citenamefont
  {Degraeve}, \citenamefont {Bury}, \citenamefont {Swerts}, \citenamefont
  {Couet}, \citenamefont {Souriau}, \citenamefont {Kundu}, \citenamefont {Rao},
  \citenamefont {Kim}, \citenamefont {Yasin}, \citenamefont {Crotti},
  \citenamefont {Linten},\ and\ \citenamefont {Kar}}]{beek_impact_2018}%
  \BibitemOpen
  \bibfield  {author} {\bibinfo {author} {\bibfnamefont {S.~V.}\ \bibnamefont
  {Beek}}, \bibinfo {author} {\bibfnamefont {B.~J.}\ \bibnamefont
  {O'Sullivan}}, \bibinfo {author} {\bibfnamefont {P.~J.}\ \bibnamefont
  {Roussel}}, \bibinfo {author} {\bibfnamefont {R.}~\bibnamefont {Degraeve}},
  \bibinfo {author} {\bibfnamefont {E.}~\bibnamefont {Bury}}, \bibinfo {author}
  {\bibfnamefont {J.}~\bibnamefont {Swerts}}, \bibinfo {author} {\bibfnamefont
  {S.}~\bibnamefont {Couet}}, \bibinfo {author} {\bibfnamefont
  {L.}~\bibnamefont {Souriau}}, \bibinfo {author} {\bibfnamefont
  {S.}~\bibnamefont {Kundu}}, \bibinfo {author} {\bibfnamefont
  {S.}~\bibnamefont {Rao}}, \bibinfo {author} {\bibfnamefont {W.}~\bibnamefont
  {Kim}}, \bibinfo {author} {\bibfnamefont {F.}~\bibnamefont {Yasin}}, \bibinfo
  {author} {\bibfnamefont {D.}~\bibnamefont {Crotti}}, \bibinfo {author}
  {\bibfnamefont {D.}~\bibnamefont {Linten}}, \ and\ \bibinfo {author}
  {\bibfnamefont {G.}~\bibnamefont {Kar}},\ }in\ \href {\doibase
  10.1109/IEDM.2018.8614617} {\emph {\bibinfo {booktitle} {2018 {IEEE}
  {International} {Electron} {Devices} {Meeting} ({IEDM})}}}\ (\bibinfo {year}
  {2018})\ pp.\ \bibinfo {pages} {25.2.1--25.2.4}\BibitemShut {NoStop}%
\bibitem [{\citenamefont {Herault}\ \emph {et~al.}(2009)\citenamefont
  {Herault}, \citenamefont {Sousa}, \citenamefont {Ducruet}, \citenamefont
  {Dieny}, \citenamefont {Conraux}, \citenamefont {Portemont}, \citenamefont
  {Mackay}, \citenamefont {Prejbeanu}, \citenamefont {Delaët}, \citenamefont
  {Cyrille},\ and\ \citenamefont {Redon}}]{herault_nanosecond_2009}%
  \BibitemOpen
  \bibfield  {author} {\bibinfo {author} {\bibfnamefont {J.}~\bibnamefont
  {Herault}}, \bibinfo {author} {\bibfnamefont {R.~C.}\ \bibnamefont {Sousa}},
  \bibinfo {author} {\bibfnamefont {C.}~\bibnamefont {Ducruet}}, \bibinfo
  {author} {\bibfnamefont {B.}~\bibnamefont {Dieny}}, \bibinfo {author}
  {\bibfnamefont {Y.}~\bibnamefont {Conraux}}, \bibinfo {author} {\bibfnamefont
  {C.}~\bibnamefont {Portemont}}, \bibinfo {author} {\bibfnamefont
  {K.}~\bibnamefont {Mackay}}, \bibinfo {author} {\bibfnamefont {I.~L.}\
  \bibnamefont {Prejbeanu}}, \bibinfo {author} {\bibfnamefont {B.}~\bibnamefont
  {Delaët}}, \bibinfo {author} {\bibfnamefont {M.~C.}\ \bibnamefont
  {Cyrille}}, \ and\ \bibinfo {author} {\bibfnamefont {O.}~\bibnamefont
  {Redon}},\ }\href {\doibase 10.1063/1.3158231} {\bibfield  {journal}
  {\bibinfo  {journal} {Journal of Applied Physics}\ }\textbf {\bibinfo
  {volume} {106}},\ \bibinfo {pages} {014505} (\bibinfo {year}
  {2009})}\BibitemShut {NoStop}%
\bibitem [{\citenamefont {Goff}\ \emph {et~al.}(2016)\citenamefont {Goff},
  \citenamefont {Nikitin},\ and\ \citenamefont
  {Devolder}}]{goff_spin-wave_2016}%
  \BibitemOpen
  \bibfield  {author} {\bibinfo {author} {\bibfnamefont {A.~L.}\ \bibnamefont
  {Goff}}, \bibinfo {author} {\bibfnamefont {V.}~\bibnamefont {Nikitin}}, \
  and\ \bibinfo {author} {\bibfnamefont {T.}~\bibnamefont {Devolder}},\ }\href
  {\doibase 10.1063/1.4953680} {\bibfield  {journal} {\bibinfo  {journal}
  {Journal of Applied Physics}\ }\textbf {\bibinfo {volume} {120}},\ \bibinfo
  {pages} {023902} (\bibinfo {year} {2016})}\BibitemShut {NoStop}%
\bibitem [{\citenamefont {Le~Goff}\ \emph {et~al.}(2014)\citenamefont
  {Le~Goff}, \citenamefont {Garcia}, \citenamefont {Vernier}, \citenamefont
  {Tahmasebi}, \citenamefont {Cornelissen}, \citenamefont {Min},\ and\
  \citenamefont {Devolder}}]{le_goff_effect_2014}%
  \BibitemOpen
  \bibfield  {author} {\bibinfo {author} {\bibfnamefont {A.}~\bibnamefont
  {Le~Goff}}, \bibinfo {author} {\bibfnamefont {K.}~\bibnamefont {Garcia}},
  \bibinfo {author} {\bibfnamefont {N.}~\bibnamefont {Vernier}}, \bibinfo
  {author} {\bibfnamefont {T.}~\bibnamefont {Tahmasebi}}, \bibinfo {author}
  {\bibfnamefont {S.}~\bibnamefont {Cornelissen}}, \bibinfo {author}
  {\bibfnamefont {T.}~\bibnamefont {Min}}, \ and\ \bibinfo {author}
  {\bibfnamefont {T.}~\bibnamefont {Devolder}},\ }\href {\doibase
  10.1109/TMAG.2014.2328664} {\bibfield  {journal} {\bibinfo  {journal} {IEEE
  Transactions on Magnetics}\ }\textbf {\bibinfo {volume} {50}},\ \bibinfo
  {pages} {1} (\bibinfo {year} {2014})}\BibitemShut {NoStop}%
\bibitem [{\citenamefont {Devolder}\ \emph
  {et~al.}(2016{\natexlab{b}})\citenamefont {Devolder}, \citenamefont {Couet},
  \citenamefont {Swerts},\ and\ \citenamefont
  {Furnemont}}]{devolder_evolution_2016}%
  \BibitemOpen
  \bibfield  {author} {\bibinfo {author} {\bibfnamefont {T.}~\bibnamefont
  {Devolder}}, \bibinfo {author} {\bibfnamefont {S.}~\bibnamefont {Couet}},
  \bibinfo {author} {\bibfnamefont {J.}~\bibnamefont {Swerts}}, \ and\ \bibinfo
  {author} {\bibfnamefont {A.}~\bibnamefont {Furnemont}},\ }\href {\doibase
  10.1063/1.4948378} {\bibfield  {journal} {\bibinfo  {journal} {Applied
  Physics Letters}\ }\textbf {\bibinfo {volume} {108}},\ \bibinfo {pages}
  {172409} (\bibinfo {year} {2016}{\natexlab{b}})}\BibitemShut {NoStop}%
\bibitem [{\citenamefont {Liu}\ \emph {et~al.}(2017)\citenamefont {Liu},
  \citenamefont {Vaysset}, \citenamefont {Swerts}, \citenamefont {Devolder},
  \citenamefont {Couet}, \citenamefont {Mertens}, \citenamefont {Lin},
  \citenamefont {Elshocht}, \citenamefont {Boeck},\ and\ \citenamefont
  {Kar}}]{liu_control_2017}%
  \BibitemOpen
  \bibfield  {author} {\bibinfo {author} {\bibfnamefont {E.}~\bibnamefont
  {Liu}}, \bibinfo {author} {\bibfnamefont {A.}~\bibnamefont {Vaysset}},
  \bibinfo {author} {\bibfnamefont {J.}~\bibnamefont {Swerts}}, \bibinfo
  {author} {\bibfnamefont {T.}~\bibnamefont {Devolder}}, \bibinfo {author}
  {\bibfnamefont {S.}~\bibnamefont {Couet}}, \bibinfo {author} {\bibfnamefont
  {S.}~\bibnamefont {Mertens}}, \bibinfo {author} {\bibfnamefont
  {T.}~\bibnamefont {Lin}}, \bibinfo {author} {\bibfnamefont {S.~V.}\
  \bibnamefont {Elshocht}}, \bibinfo {author} {\bibfnamefont {J.~D.}\
  \bibnamefont {Boeck}}, \ and\ \bibinfo {author} {\bibfnamefont
  {G.}~\bibnamefont {Kar}},\ }\href {\doibase 10.1109/TMAG.2017.2701553}
  {\bibfield  {journal} {\bibinfo  {journal} {IEEE Transactions on Magnetics}\
  }\textbf {\bibinfo {volume} {PP}},\ \bibinfo {pages} {1} (\bibinfo {year}
  {2017})}\BibitemShut {NoStop}%
\bibitem [{\citenamefont {Devolder}\ \emph
  {et~al.}(2019{\natexlab{b}})\citenamefont {Devolder}, \citenamefont
  {Carpenter}, \citenamefont {Rao}, \citenamefont {Kim}, \citenamefont {Couet},
  \citenamefont {Swerts},\ and\ \citenamefont {Kar}}]{devolder_offset_2019}%
  \BibitemOpen
  \bibfield  {author} {\bibinfo {author} {\bibfnamefont {T.}~\bibnamefont
  {Devolder}}, \bibinfo {author} {\bibfnamefont {R.}~\bibnamefont {Carpenter}},
  \bibinfo {author} {\bibfnamefont {S.}~\bibnamefont {Rao}}, \bibinfo {author}
  {\bibfnamefont {W.}~\bibnamefont {Kim}}, \bibinfo {author} {\bibfnamefont
  {S.}~\bibnamefont {Couet}}, \bibinfo {author} {\bibfnamefont
  {J.}~\bibnamefont {Swerts}}, \ and\ \bibinfo {author} {\bibfnamefont {G.~S.}\
  \bibnamefont {Kar}},\ }\href {\doibase 10.1088/1361-6463/ab1b07} {\bibfield
  {journal} {\bibinfo  {journal} {Journal of Physics D: Applied Physics}\
  }\textbf {\bibinfo {volume} {52}},\ \bibinfo {pages} {274001} (\bibinfo
  {year} {2019}{\natexlab{b}})}\BibitemShut {NoStop}%
\bibitem [{\citenamefont {Cuchet}\ \emph {et~al.}(2014)\citenamefont {Cuchet},
  \citenamefont {Rodmacq}, \citenamefont {Auffret}, \citenamefont {Sousa},\
  and\ \citenamefont {Dieny}}]{cuchet_influence_2014}%
  \BibitemOpen
  \bibfield  {author} {\bibinfo {author} {\bibfnamefont {L.}~\bibnamefont
  {Cuchet}}, \bibinfo {author} {\bibfnamefont {B.}~\bibnamefont {Rodmacq}},
  \bibinfo {author} {\bibfnamefont {S.}~\bibnamefont {Auffret}}, \bibinfo
  {author} {\bibfnamefont {R.~C.}\ \bibnamefont {Sousa}}, \ and\ \bibinfo
  {author} {\bibfnamefont {B.}~\bibnamefont {Dieny}},\ }\href {\doibase
  10.1063/1.4892450} {\bibfield  {journal} {\bibinfo  {journal} {Applied
  Physics Letters}\ }\textbf {\bibinfo {volume} {105}},\ \bibinfo {pages}
  {052408} (\bibinfo {year} {2014})}\BibitemShut {NoStop}%
\bibitem [{\citenamefont {Worledge}\ \emph {et~al.}(2011)\citenamefont
  {Worledge}, \citenamefont {Hu}, \citenamefont {Abraham}, \citenamefont {Sun},
  \citenamefont {Trouilloud}, \citenamefont {Nowak}, \citenamefont {Brown},
  \citenamefont {Gaidis}, \citenamefont {O’Sullivan},\ and\ \citenamefont
  {Robertazzi}}]{worledge_spin_2011}%
  \BibitemOpen
  \bibfield  {author} {\bibinfo {author} {\bibfnamefont {D.~C.}\ \bibnamefont
  {Worledge}}, \bibinfo {author} {\bibfnamefont {G.}~\bibnamefont {Hu}},
  \bibinfo {author} {\bibfnamefont {D.~W.}\ \bibnamefont {Abraham}}, \bibinfo
  {author} {\bibfnamefont {J.~Z.}\ \bibnamefont {Sun}}, \bibinfo {author}
  {\bibfnamefont {P.~L.}\ \bibnamefont {Trouilloud}}, \bibinfo {author}
  {\bibfnamefont {J.}~\bibnamefont {Nowak}}, \bibinfo {author} {\bibfnamefont
  {S.}~\bibnamefont {Brown}}, \bibinfo {author} {\bibfnamefont {M.~C.}\
  \bibnamefont {Gaidis}}, \bibinfo {author} {\bibfnamefont {E.~J.}\
  \bibnamefont {O’Sullivan}}, \ and\ \bibinfo {author} {\bibfnamefont
  {R.~P.}\ \bibnamefont {Robertazzi}},\ }\href {\doibase 10.1063/1.3536482}
  {\bibfield  {journal} {\bibinfo  {journal} {Applied Physics Letters}\
  }\textbf {\bibinfo {volume} {98}},\ \bibinfo {pages} {022501} (\bibinfo
  {year} {2011})}\BibitemShut {NoStop}%
\bibitem [{\citenamefont {Gajek}\ \emph {et~al.}(2012)\citenamefont {Gajek},
  \citenamefont {Nowak}, \citenamefont {Sun}, \citenamefont {Trouilloud},
  \citenamefont {O’Sullivan}, \citenamefont {Abraham}, \citenamefont
  {Gaidis}, \citenamefont {Hu}, \citenamefont {Brown}, \citenamefont {Zhu},
  \citenamefont {Robertazzi}, \citenamefont {Gallagher},\ and\ \citenamefont
  {Worledge}}]{gajek_spin_2012}%
  \BibitemOpen
  \bibfield  {author} {\bibinfo {author} {\bibfnamefont {M.}~\bibnamefont
  {Gajek}}, \bibinfo {author} {\bibfnamefont {J.~J.}\ \bibnamefont {Nowak}},
  \bibinfo {author} {\bibfnamefont {J.~Z.}\ \bibnamefont {Sun}}, \bibinfo
  {author} {\bibfnamefont {P.~L.}\ \bibnamefont {Trouilloud}}, \bibinfo
  {author} {\bibfnamefont {E.~J.}\ \bibnamefont {O’Sullivan}}, \bibinfo
  {author} {\bibfnamefont {D.~W.}\ \bibnamefont {Abraham}}, \bibinfo {author}
  {\bibfnamefont {M.~C.}\ \bibnamefont {Gaidis}}, \bibinfo {author}
  {\bibfnamefont {G.}~\bibnamefont {Hu}}, \bibinfo {author} {\bibfnamefont
  {S.}~\bibnamefont {Brown}}, \bibinfo {author} {\bibfnamefont
  {Y.}~\bibnamefont {Zhu}}, \bibinfo {author} {\bibfnamefont {R.~P.}\
  \bibnamefont {Robertazzi}}, \bibinfo {author} {\bibfnamefont {W.~J.}\
  \bibnamefont {Gallagher}}, \ and\ \bibinfo {author} {\bibfnamefont {D.~C.}\
  \bibnamefont {Worledge}},\ }\href {\doibase 10.1063/1.3694270} {\bibfield
  {journal} {\bibinfo  {journal} {Applied Physics Letters}\ }\textbf {\bibinfo
  {volume} {100}},\ \bibinfo {pages} {132408} (\bibinfo {year}
  {2012})}\BibitemShut {NoStop}%
\bibitem [{\citenamefont {Tomczak}\ \emph {et~al.}(2016)\citenamefont
  {Tomczak}, \citenamefont {Swerts}, \citenamefont {Mertens}, \citenamefont
  {Lin}, \citenamefont {Couet}, \citenamefont {Liu}, \citenamefont {Sankaran},
  \citenamefont {Pourtois}, \citenamefont {Kim}, \citenamefont {Souriau},
  \citenamefont {Elshocht}, \citenamefont {Kar},\ and\ \citenamefont
  {Furnemont}}]{tomczak_thin_2016}%
  \BibitemOpen
  \bibfield  {author} {\bibinfo {author} {\bibfnamefont {Y.}~\bibnamefont
  {Tomczak}}, \bibinfo {author} {\bibfnamefont {J.}~\bibnamefont {Swerts}},
  \bibinfo {author} {\bibfnamefont {S.}~\bibnamefont {Mertens}}, \bibinfo
  {author} {\bibfnamefont {T.}~\bibnamefont {Lin}}, \bibinfo {author}
  {\bibfnamefont {S.}~\bibnamefont {Couet}}, \bibinfo {author} {\bibfnamefont
  {E.}~\bibnamefont {Liu}}, \bibinfo {author} {\bibfnamefont {K.}~\bibnamefont
  {Sankaran}}, \bibinfo {author} {\bibfnamefont {G.}~\bibnamefont {Pourtois}},
  \bibinfo {author} {\bibfnamefont {W.}~\bibnamefont {Kim}}, \bibinfo {author}
  {\bibfnamefont {L.}~\bibnamefont {Souriau}}, \bibinfo {author} {\bibfnamefont
  {S.~V.}\ \bibnamefont {Elshocht}}, \bibinfo {author} {\bibfnamefont
  {G.}~\bibnamefont {Kar}}, \ and\ \bibinfo {author} {\bibfnamefont
  {A.}~\bibnamefont {Furnemont}},\ }\href {\doibase 10.1063/1.4940772}
  {\bibfield  {journal} {\bibinfo  {journal} {Applied Physics Letters}\
  }\textbf {\bibinfo {volume} {108}},\ \bibinfo {pages} {042402} (\bibinfo
  {year} {2016})}\BibitemShut {NoStop}%
\bibitem [{\citenamefont {Couet}\ \emph {et~al.}(2017)\citenamefont {Couet},
  \citenamefont {Devolder}, \citenamefont {Swerts}, \citenamefont {Mertens},
  \citenamefont {Lin}, \citenamefont {Liu}, \citenamefont {Van~Elshocht},\ and\
  \citenamefont {Sankar~Kar}}]{couet_impact_2017}%
  \BibitemOpen
  \bibfield  {author} {\bibinfo {author} {\bibfnamefont {S.}~\bibnamefont
  {Couet}}, \bibinfo {author} {\bibfnamefont {T.}~\bibnamefont {Devolder}},
  \bibinfo {author} {\bibfnamefont {J.}~\bibnamefont {Swerts}}, \bibinfo
  {author} {\bibfnamefont {S.}~\bibnamefont {Mertens}}, \bibinfo {author}
  {\bibfnamefont {T.}~\bibnamefont {Lin}}, \bibinfo {author} {\bibfnamefont
  {E.}~\bibnamefont {Liu}}, \bibinfo {author} {\bibfnamefont {S.}~\bibnamefont
  {Van~Elshocht}}, \ and\ \bibinfo {author} {\bibfnamefont {G.}~\bibnamefont
  {Sankar~Kar}},\ }\href {\doibase 10.1063/1.5000992} {\bibfield  {journal}
  {\bibinfo  {journal} {Applied Physics Letters}\ }\textbf {\bibinfo {volume}
  {111}},\ \bibinfo {pages} {152406} (\bibinfo {year} {2017})}\BibitemShut
  {NoStop}%
\bibitem [{\citenamefont {Gottwald}\ \emph {et~al.}(2013)\citenamefont
  {Gottwald}, \citenamefont {Kan}, \citenamefont {Lee}, \citenamefont {Kang},\
  and\ \citenamefont {Fullerton}}]{gottwald_paramagnetic_2013}%
  \BibitemOpen
  \bibfield  {author} {\bibinfo {author} {\bibfnamefont {M.}~\bibnamefont
  {Gottwald}}, \bibinfo {author} {\bibfnamefont {J.~J.}\ \bibnamefont {Kan}},
  \bibinfo {author} {\bibfnamefont {K.}~\bibnamefont {Lee}}, \bibinfo {author}
  {\bibfnamefont {S.~H.}\ \bibnamefont {Kang}}, \ and\ \bibinfo {author}
  {\bibfnamefont {E.~E.}\ \bibnamefont {Fullerton}},\ }\href {\doibase
  10.1063/1.4817897} {\bibfield  {journal} {\bibinfo  {journal} {APL
  Materials}\ }\textbf {\bibinfo {volume} {1}},\ \bibinfo {pages} {022102}
  (\bibinfo {year} {2013})}\BibitemShut {NoStop}%
\bibitem [{\citenamefont {Kim}\ \emph {et~al.}(2015)\citenamefont {Kim},
  \citenamefont {Lee}, \citenamefont {An}, \citenamefont {Yang}, \citenamefont
  {Chung}, \citenamefont {Park},\ and\ \citenamefont
  {Hong}}]{kim_ultrathin_2015}%
  \BibitemOpen
  \bibfield  {author} {\bibinfo {author} {\bibfnamefont {J.-H.}\ \bibnamefont
  {Kim}}, \bibinfo {author} {\bibfnamefont {J.-B.}\ \bibnamefont {Lee}},
  \bibinfo {author} {\bibfnamefont {G.-G.}\ \bibnamefont {An}}, \bibinfo
  {author} {\bibfnamefont {S.-M.}\ \bibnamefont {Yang}}, \bibinfo {author}
  {\bibfnamefont {W.-S.}\ \bibnamefont {Chung}}, \bibinfo {author}
  {\bibfnamefont {H.-S.}\ \bibnamefont {Park}}, \ and\ \bibinfo {author}
  {\bibfnamefont {J.-P.}\ \bibnamefont {Hong}},\ }\href {\doibase
  10.1038/srep16903} {\bibfield  {journal} {\bibinfo  {journal} {Scientific
  Reports}\ }\textbf {\bibinfo {volume} {5}},\ \bibinfo {pages} {16903}
  (\bibinfo {year} {2015})}\BibitemShut {NoStop}%
\bibitem [{\citenamefont {Liu}\ \emph {et~al.}(2016)\citenamefont {Liu},
  \citenamefont {Yu}, \citenamefont {Zhu}, \citenamefont {Zhong}, \citenamefont
  {Khamis},\ and\ \citenamefont {Zhu}}]{liu_high_2016}%
  \BibitemOpen
  \bibfield  {author} {\bibinfo {author} {\bibfnamefont {Y.}~\bibnamefont
  {Liu}}, \bibinfo {author} {\bibfnamefont {T.}~\bibnamefont {Yu}}, \bibinfo
  {author} {\bibfnamefont {Z.}~\bibnamefont {Zhu}}, \bibinfo {author}
  {\bibfnamefont {H.}~\bibnamefont {Zhong}}, \bibinfo {author} {\bibfnamefont
  {K.~M.}\ \bibnamefont {Khamis}}, \ and\ \bibinfo {author} {\bibfnamefont
  {K.}~\bibnamefont {Zhu}},\ }\href {\doibase 10.1016/j.jmmm.2016.02.099}
  {\bibfield  {journal} {\bibinfo  {journal} {Journal of Magnetism and Magnetic
  Materials}\ }\textbf {\bibinfo {volume} {410}},\ \bibinfo {pages} {123}
  (\bibinfo {year} {2016})}\BibitemShut {NoStop}%
\bibitem [{\citenamefont {Devolder}\ \emph
  {et~al.}(2016{\natexlab{c}})\citenamefont {Devolder}, \citenamefont
  {Le~Goff},\ and\ \citenamefont {Nikitin}}]{devolder_size_2016}%
  \BibitemOpen
  \bibfield  {author} {\bibinfo {author} {\bibfnamefont {T.}~\bibnamefont
  {Devolder}}, \bibinfo {author} {\bibfnamefont {A.}~\bibnamefont {Le~Goff}}, \
  and\ \bibinfo {author} {\bibfnamefont {V.}~\bibnamefont {Nikitin}},\ }\href
  {\doibase 10.1103/PhysRevB.93.224432} {\bibfield  {journal} {\bibinfo
  {journal} {Physical Review B}\ }\textbf {\bibinfo {volume} {93}},\ \bibinfo
  {pages} {224432} (\bibinfo {year} {2016}{\natexlab{c}})}\BibitemShut
  {NoStop}%
\bibitem [{\citenamefont {Slonczewski}(2005)}]{slonczewski_currents_2005}%
  \BibitemOpen
  \bibfield  {author} {\bibinfo {author} {\bibfnamefont {J.~C.}\ \bibnamefont
  {Slonczewski}},\ }\href {\doibase 10.1103/PhysRevB.71.024411} {\bibfield
  {journal} {\bibinfo  {journal} {Physical Review B}\ }\textbf {\bibinfo
  {volume} {71}},\ \bibinfo {pages} {024411} (\bibinfo {year}
  {2005})}\BibitemShut {NoStop}%
\end{thebibliography}
%

\end{document}